# Role of Polarization-Photon Coupling in Ultrafast Terahertz Excitation of Ferroelectrics

Shihao Zhuang and Jia-Mian Hu[*]

*Department of Materials Science and Engineering, University of Wisconsin-Madison, Madison, WI, 53706, USA*

**Abstract**

We investigate the role of polarization-photon coupling (specifically, polarization-oscillation-induced radiation electric field) in the excitation of ferroelectric thin films by an ultrafast terahertz (THz) electric-field pulse. Analytical theory is developed to predict how the frequencies and relaxation time of three-dimensional soft mode phonons (intrinsic polarization oscillation) are modulated by radiation electric field and epitaxial strain. Ultrafast THz-pulse-driven excitation of harmonic polarization oscillation in strained single-domain ferroelectric thin film is then simulated using a dynamical phase-field model that considers the coupled strain-polarization-photon dynamics. The frequencies and relaxational time extracted from such numerical simulations agree well with analytical predictions. In relatively thin films, it is predicted that the radiation electric field slightly reduces the frequencies but significantly shortens the relaxational time. These results demonstrate the necessity of considering polarization-photon coupling in understanding and predicting the response of ferroelectric materials to ultrafast pulses of THz and higher frequencies.

*E-mail: jhu238@wisc.edu

# I. Introduction

Ultrafast light-matter interaction is a promising route to understanding and controlling quantum materials where the four fundamental degrees of freedoms (charge, spin, orbit, and lattice) are dynamically intertwined [1,2]. By tailoring the wavelength, duration, and amplitude of ultrafast light pulse, one can probe the coupling among different quasiparticles (e.g., phonons, magnons), switch the material to a degenerate ground-state or a hidden metastable state of matter, or induce phase transitions [3,4]. Ferroelectric materials have a spontaneous polarization that can be switched by its conjugate electric field or other non-conjugate stimuli. Ultrafast light control of ferroelectrics offers potential for developing photoferroic devices with order of magnitude higher operation speed than those controlled by voltage [5]. In archetypal ferroelectric system such as $PbTiO_3$ and $BaTiO_3$, the development of spontaneous polarization **P** is attributed to the condensation of a particular soft mode phonon $A_1(TO)$ (see Fig. 1a) below the Curie temperature [6–8]. Depending on wavelength of the light pulse, ultrafast light control of polarization in ferroelectrics can occur through various nonthermal mechanisms. These include the nonlinear coupling between the soft mode phonons and the electric-field of the near-infrared pluses via impulsive stimulated Raman Scattering [9,10], linear coupling between soft mode phonon and the electric field of terahertz (THz) pulse [11–19], indirect manipulation of soft mode phonons by mid-infrared pulses [20–25], and above-bandgap photo-carrier excitation [26–29].

Among these mechanisms, the use of THz electric-field pulse to drive soft mode phonons is most straightforward [5]. Since the proposal for achieving picosecond (*ps*) polarization reversal in single-domain $PbTiO_3$ by shaped THz electric-field pulses [11], ultrafast THz-field induced *ps* polarization modulation were experimentally observed in various ferroelectric materials systems [12,14–17,19]. Moreover, a recent experiment shows that THz electric-field pulse can transiently drive the quantum paraelectric $SrTiO_3$ into a ferroelectric state [16]. Furthermore, molecular dynamics simulations predict that ultrafast THz electric-field pulse can uncover hidden phase of polarization order in relaxor ferroelectric systems [18]. Despite these extensive works, the role of bidirectional polarization-photon coupling, i.e., the emission of electromagnetic (EM) waves from the oscillating polarization and the backaction of the emitted EM waves on polarization dynamics, has not yet been addressed. Experimental reports on time-resolved THz pump-THz probe measurement of ferroelectrics remain scarce. Computationally, the polarization-photon coupling has not yet been considered in existing atomistic [11,13,18] and mesoscale [19,30] simulations, which is partly due to the high computational burden of modeling the generation and propagation of long-wavelength (millimeter-scale) EM waves in nanometer-scale systems.

In this article, we computationally demonstrated the necessity of considering bidirectional polarization-photon coupling in understanding and predicting ultrafast THz excitation of ferroelectrics. This paper is organized as follows. In Sect. II, we analytically calculate the frequencies and relaxation time of the soft mode phonon by considering a harmonic polarization oscillation in single-domain ferroelectrics. Influences of polarization-oscillation-induced radiation electric field on the frequencies and relaxation time of the soft mode phonons are analytically predicted. Effects of epitaxial strain in ferroelectric thin films on the soft mode phonon frequencies are also calculated. In Sect. III, we numerically simulate the excitation of a single-domain, in-plane-polarized $(100)_{pc}$ $BaTiO_3$ thin film (pc: pseudocubic) by applying an ultrafast THz electric-field pulse. This is achieved using an in-house dynamical phase-field model that considers coupled dynamics of strain, polarization, and EM waves, where Graphics Processing Unit (GPU) acceleration was performed to mitigate the high computational cost. The simulations results confirmed the analytical predictions. Since our analytical and numerical models are independent from each other, the consistency in the prediction results simultaneously demonstrates the physical validity of our analytical theory and high numerical accuracy of our dynamical phase-field model. The paper is concluded with discussion in Sect. IV.

**II. Analytical theory for the frequencies and relaxation time of soft mode phonons**

In this section, we derive analytical expressions for the frequencies of three-dimensional soft mode phonons, i.e., the intrinsic oscillation frequencies of the $P_x$, $P_y$, $P_z$ of the spontaneous polarization **P**, as well as their relaxation time. Our analytical formulations are developed partly based on the work by Morozovska and colleagues [31], but considers the influence of both the substrate-induced strain (via lattice/thermal mismatch) in an epitaxial ferroelectric film and the polarization-oscillation-induced radiation electric field.

As an example, let us consider a $(100)_{pc}$ BaTiO$_3$ film epitaxially grown on a $(110)_O$ PrScO$_3$ substrate (O: orthorhombic), as shown in Fig. 1b. In such heterostructure, the combination of anisotropic mismatch strain and interfacial symmetry breaking can stabilize an in-plane-polarized single ferroelectric domain in the BaTiO$_3$ film, which has been demonstrated experimentally [32]. Figure 1c shows the landscape of the total free energy density of such anisotropically strained $(100)_{pc}$ BaTiO$_3$ thin film at thermodynamic equilibrium, where the anisotropy mismatch strain, $\varepsilon_{xx}^{mis} = 0.5\%$, $\varepsilon_{yy}^{mis} = 0.01\%$ and $\varepsilon_{xy}^{mis} = 0$ [32]. The total free energy density in Fig. 1c, which includes the Landau free energy and elastic energy densities, is calculated as a function of the $P_x$ and $P_y$. As shown, the equilibrium polarizations are located at the two degenerate energy minima at $\pm \mathbf{P}^0$, where $\mathbf{P}^0 = (P_x^0, P_y^0, P_z^0) = (0.294, 0, 0)$ C m$^{-2}$. We note that $P_z$ is always 0 at equilibrium because of the depolarization field and because the anisotropic mismatch strain favors an in-plane polarization along the $x$-axis. An ultrafast external stimulus such as a THz light pulse can perturb the polarization away from the $\mathbf{P}^0$. After then, the polarization will oscillate around the $\mathbf{P}^0$ at its intrinsic frequency (i.e., the frequency of soft mode phonon) until it returns to equilibrium via damping. A larger slope of the local free energy landscape will lead to a higher intrinsic oscillation frequency. Figure 1d shows the free energy landscape near the $\mathbf{P}^0$. As shown, the slope of the line AB (along which $P_y$=0 and $P_x$ varies from 0.25 to 0.35 C m$^{-2}$) is larger than the slope of the line CD (along which $P_x$=0.294 C m$^{-2}$ and $P_y$ varies from -0.2 to 0.2 C m$^{-2}$). Therefore, the intrinsic oscillation frequency of the $P_x$ should be larger than that of the $P_y$. For clearer comparison, Figure 1e shows the data along the line AB and CD in a two-dimensional plot within a smaller range of variation (below $\pm 0.05$ C m$^{-2}$). It is noteworthy that such large discrepancy in slope mainly results from the landscape of the Landau free energy density rather than the anisotropic mismatch strain. For demonstration, the line profiles of the Landau free energy density along $P_x$ and $P_y$ near the $\mathbf{P}^0$ are also plotted in Fig. 1e. As seen, the slope for the variation of $P_x$ is still larger than that for the $P_y$. Comparing the solid and dashed lines, it can be seen that the anisotropic mismatch strain increases the slope for both the variation of $P_x$ and $P_y$, indicating strain-induced enhancement in their oscillation frequencies. Note that the $\mathbf{P}^0 = (P_x^0, P_y^0, P_z^0) = (0.26, 0, 0)$ C m$^{-2}$ under zero strain.

To derive the frequency and relaxation time of the soft mode phonons, we start by considering a harmonic polarization oscillation in single-domain ferroelectric materials and write the equation of motion of the polarization, given as,

$$\mu \frac{\partial^2 P_i}{\partial t^2} + \gamma \frac{\partial P_i}{\partial t} = E_i^{\text{eff}} \quad (1)$$

where $P_i$ ($i = x, y, z$) are the three components of the spontaneous polarization in Cartesian coordinates; $\mu$ and $\gamma$ are the mass and damping coefficient of the polarization dynamics, respectively; $E_i^{\text{eff}} = -\frac{\partial f^{\text{tot}}}{\partial P_i}$ is the total effective electric field, where $f^{\text{tot}}$ is the total free energy density. For a single-domain strained ferroelectric thin film, $f^{\text{tot}} = f^{\text{Landau}} + f^{\text{elas}} + f^{\text{elec}}$ is a sum of the Landau, elastic, and electrical free energy density, respectively. For BaTiO$_3$, an eighth-order Landau free energy density is utilized, and its expression can be found in ref. [33]. The corresponding effective field $E_i^{\text{Landau}}$ is calculated as $-\frac{\partial f^{\text{Landau}}}{\partial P_i}$,

$$E_i^{\text{Landau}} = -2\alpha_1 P_i$$
$$-4\alpha_{11}P_i^3 - 2\alpha_{12}P_i(P_j^2 + P_k^2)$$
$$-6\alpha_{111}P_i^5 - 4\alpha_{112}P_i^3(P_j^2 + P_k^2) - 2\alpha_{112}P_i(P_j^4 + P_k^4) - 2\alpha_{123}P_iP_j^2P_k^2$$
$$-8\alpha_{1111}P_i^7 - 2\alpha_{1112}P_i[P_j^6 + P_k^6 + 3P_i^4(P_j^2 + P_k^2)] - 4\alpha_{1122}P_i^3(P_j^4 + P_k^4) - 2\alpha_{1123}P_iP_j^2P_k^2(2P_i^2 + P_j^2 + P_k^2), \quad (2)$$

where $i = x, y, z$, and $j \neq i, k \neq i, j$; $\boldsymbol{\alpha}$ are Landau–Devonshire coefficients among which $\alpha_1 = \alpha_0(T - T_c)$ is a linear function of temperature, and $T_c$ is the Curie temperature. The values of these coefficients are also provided in ref. [33].

The $E_i^{\text{elas}}$ is contributed by the elastic energy $f^{\text{elas}}$ and calculated as $-\frac{\partial f^{\text{elas}}}{\partial P_i}$, and the expression of $f^{\text{elas}}$ can be found in ref. [34],

$$E_i^{\text{elas}} = [2q_{11}(\varepsilon_{ii} - \varepsilon_{ii}^0) + 2q_{12}(\varepsilon_{jj} + \varepsilon_{kk} - \varepsilon_{jj}^0 - \varepsilon_{kk}^0)]P_i + 2q_{44}[(\varepsilon_{ij} - \varepsilon_{ij}^0)P_j + (\varepsilon_{ik} - \varepsilon_{ik}^0)P_k]. \quad (3)$$

Here $q_{11} = c_{11}Q_{11} + 2c_{12}Q_{12}$, $q_{12} = c_{11}Q_{12} + c_{12}(Q_{11} + Q_{12})$ and $q_{44} = 2c_{44}Q_{44}$, where $\mathbf{c}$ and $\mathbf{Q}$ are the elastic stiffness and electrostrictive tensor, respectively [34]; $\boldsymbol{\varepsilon}$ is total strain and $\boldsymbol{\varepsilon}^0$ is stress-free strain caused by the polarization via electrostriction, given by,

$$\varepsilon_{ii}^0 = Q_{11}P_i^2 + Q_{12}(P_j^2 + P_k^2) \tag{4a}$$

$$\varepsilon_{ij}^0 = Q_{44}P_iP_j \tag{4b}$$

with $i = x, y, z$, and $j \neq i, k \neq i, j$. For a strained single-domain ferroelectric film, the total strain $\boldsymbol{\varepsilon}(\mathbf{P}^0)$ at the initial equilibrium state can be analytically obtained by solving the mechanical equilibrium equation $\frac{\partial \sigma_{iz}}{\partial z} = 0$ ($i = x, y, z$) with the stress $\sigma_{ij} = c_{ijkl}(\varepsilon_{kl} - \varepsilon_{kl}^0)$. After some algebra, one has,

$$\varepsilon_{zz}(\mathbf{P}^0) = -\frac{c_{12}}{c_{11}}(\varepsilon_{xx}^{\text{mis}} + \varepsilon_{yy}^{\text{mis}}) + \varepsilon_{zz}^0(\mathbf{P}^0), \tag{5a}$$

$$\varepsilon_{yz}(\mathbf{P}^0) = \varepsilon_{yz}^0(\mathbf{P}^0), \tag{5b}$$

$$\varepsilon_{xz}(\mathbf{P}^0) = \varepsilon_{xz}^0(\mathbf{P}^0). \tag{5c}$$

Here $\boldsymbol{\varepsilon}^{\text{mis}}$ is the mismatch strain in the ferroelectric film induced by the substrate, and,

$$\varepsilon_{xx}(\mathbf{P}^0) = \varepsilon_{xx}^{\text{mis}} + \varepsilon_{xx}^0(\mathbf{P}^0), \tag{5d}$$

$$\varepsilon_{yy}(\mathbf{P}^0) = \varepsilon_{yy}^{\text{mis}} + \varepsilon_{yy}^0(\mathbf{P}^0), \tag{5e}$$

$$\varepsilon_{xy}(\mathbf{P}^0) = \varepsilon_{xy}^{\text{mis}} + \varepsilon_{xy}^0(\mathbf{P}^0). \tag{5f}$$

Strictly speaking, the total strain in Eq. (3) should also contain a temporal variation component, $\Delta\varepsilon(t)$, which is produced by the polarization oscillation $\Delta\mathbf{P}(t)$ via electrostriction. However, the magnitude of such secondary strain would be negligible (~$10^{-5}$) for a harmonic polarization oscillation with relatively small $\Delta\mathbf{P}(t)$ (see Supplementary Material 1).

The electrical energy density $f^{\mathrm{elec}} = -(\mathbf{E}^{\mathrm{ext}}+\mathbf{E}^{\mathrm{d}})\cdot\mathbf{P}$, where the external electric field $\mathbf{E}^{\mathrm{ext}}=\mathbf{E}^{\mathrm{app}}+\mathbf{E}^{\mathrm{EM}}$ is the sum of the applied electric field and the radiation electric field inside the ferroelectric thin film. $\mathbf{E}^{\mathrm{EM}}$ describes the backaction of the emitted EM wave on polarization dynamics. For a film that is infinitely large in the $xy$ plane, $\mathbf{E}^{\mathrm{EM}}$ propagates along the film thickness direction in the form of a plane wave with $E_z^{\mathrm{EM}} = 0$. The magnitudes of $E_x^{\mathrm{EM}}$ and $E_y^{\mathrm{EM}}$ are proportional to $\frac{\partial \mathbf{P}}{\partial t}$. When the film is thin enough ($d \ll v/f$), one has $E_i^{\mathrm{EM}}(t) \approx -\frac{1}{2}\sqrt{\frac{\mu_0}{\kappa_0}}d\frac{\partial P_i}{\partial t}(t)$ over the entire film thickness, where $f$ and $v$ are the frequency and propagation speed of the EM wave in the ferroelectric film, respectively (see detailed derivation in Supplementary Material 2). The depolarizing field $\mathbf{E}^{\mathrm{d}}$ satisfies the Gauss's law $\nabla\cdot\mathbf{D} = \rho_{\mathrm{f}}$, where $\mathbf{D} = \kappa_0\kappa_{\mathrm{b}}\mathbf{E}^{\mathrm{d}}+\mathbf{P}$ is electric displacement field; $\kappa_0$ and $\kappa_{\mathrm{b}}$ are vacuum permittivity and background dielectric constant respectively; $\rho_{\mathrm{f}}$ is the free charge density. For a thin film with a spatially uniform $\mathbf{P}$, an infinitely large $xy$ plane, and $\rho_{\mathrm{f}} = 0$, $\mathbf{E}^{\mathrm{d}}$ can be solved analytically as $(E_x^{\mathrm{d}}, E_y^{\mathrm{d}}, E_z^{\mathrm{d}}) = (0, 0, -\frac{P_z}{\kappa_0\kappa_{\mathrm{b}}})$.

We have now derived a complete analytical expression of the $E_i^{\mathrm{eff}}$ in Eq. (1), with $E_i^{\mathrm{eff}} = E_i^{\mathrm{Landau}}+E_i^{\mathrm{elas}}+E_i^{\mathrm{app}}+E_i^{\mathrm{EM}}+E_i^{\mathrm{d}}$. Next, we linearize Eq. (1) by assuming that $P_i=P_i^0+\Delta P_i(t)$, where the temporal polarization oscillation $\Delta P_i(t)=\Delta P_i^0 e^{i\omega t}e^{-\lambda t}$ takes the form of a damped perturbation; $\omega$ and $\Delta P_i^0$ are the angular frequency and peak amplitude of the oscillation, respectively; $\lambda$ is an auxiliary damping coefficient. Based on these, Eq. (1) can be linearized into,

$$\left(\mu\lambda^2-\mu\omega^2-\gamma\lambda\right)\Delta P_i(t)+i(\gamma\omega-2\omega\lambda\mu)\Delta P_i(t)=E_i^{\mathrm{eff}} \qquad (6)$$

Regarding the effective fields on the right-hand side, the depolarization field can be written as $\mathbf{E}^{\mathrm{d}} = \mathbf{E}^{\mathrm{d}}(\mathbf{P}^0)+\Delta\mathbf{E}^{\mathrm{d}}(t)$, where $\mathbf{E}^{\mathrm{d}}(\mathbf{P}^0)=(0, 0, -\frac{P_z^0}{\kappa_0\kappa_{\mathrm{b}}})$ is the field at the initial equilibrium state and $\Delta\mathbf{E}^{\mathrm{d}}(t)=(0, 0, -\frac{\Delta P_z}{\kappa_0\kappa_{\mathrm{b}}})$ describes its temporal variation. The radiating electric field can be rewritten as $E_i^{\mathrm{EM}}(t) \approx -\frac{1}{2}\sqrt{\frac{\mu_0}{\kappa_0}}d(i\omega-\lambda)\Delta P_i(t)$ ($i = x, y$). To extract the terms of different orders of $\Delta P_i(t)$ from the $E_i^{\mathrm{Landau}}$ and $E_i^{\mathrm{elas}}$, we perform Taylor expansion for the function of these two effective fields near the equilibrium point $\mathbf{P}^0$ using $\Delta P_i$ as the only variable,

$$E_i^{\mathrm{Landau}}=E_i^{\mathrm{Landau}}(\mathbf{P}^0)+\frac{\partial E_i^{\mathrm{Landau}}}{\partial P_i}(\mathbf{P}^0)\times\Delta P_i+\frac{1}{2!}\frac{\partial^2 E_i^{\mathrm{Landau}}}{\partial P_i^2}(\mathbf{P}^0)\times\Delta P_i^2+\cdots \qquad (7)$$

$$E_i^{\mathrm{elas}}=E_i^{\mathrm{elas}}(\mathbf{P}^0)+\frac{\partial E_i^{\mathrm{elas}}}{\partial P_i}(\mathbf{P}^0)\times\Delta P_i+\frac{1}{2!}\frac{\partial^2 E_i^{\mathrm{elas}}}{\partial P_i^2}(\mathbf{P}^0)\times\Delta P_i^2+\cdots \qquad (8)$$

where the third and higher order terms are not shown for simplicity. For near-equilibrium excitation, $\Delta P_i^2$ and higher orders are negligibly small and can be omitted. As a result, the linearized equation of motion for polarization (Eq. 6) can be further rewritten as

$$\left(\mu\lambda^2-\mu\omega^2-\gamma\lambda\right)+i\left(\gamma\omega-2\omega\lambda\mu+\frac{1}{2}\sqrt{\frac{\mu_0}{\kappa_0}}d\omega\right)=\frac{\partial E_i^{\mathrm{Landau}}}{\partial P_i}(\mathbf{P}^0)+\frac{\partial E_i^{\mathrm{elas}}}{\partial P_i}(\mathbf{P}^0)+\frac{1}{2}\sqrt{\frac{\mu_0}{\kappa_0}}d\lambda, (i=x,y) \quad (9a)$$

$$\left(\mu\lambda^2-\mu\omega^2-\gamma\lambda\right)+i(\gamma\omega-2\omega\lambda\mu)=\frac{\partial E_z^{\mathrm{Landau}}}{\partial P_z}(\mathbf{P}^0)+\frac{\partial E_z^{\mathrm{elas}}}{\partial P_z}(\mathbf{P}^0)-\frac{1}{\kappa_0\kappa_b}. \quad (9b)$$

Note that $E_i^{\mathrm{Landau}}(\mathbf{P}^0) + E_i^{\mathrm{elas}}(\mathbf{P}^0) + E_i^{\mathrm{d}}(\mathbf{P}^0) + E_i^{\mathrm{app}}$, which refers to the total effective field at initial equilibrium state, equals to 0. By letting the imaginary terms on the left-hand side of Eqs. (9a-b) be 0, the expression for the auxiliary damping coefficient $\lambda$ can be derived as,

$$\lambda_i = \left(\gamma + \frac{1}{2}\sqrt{\frac{\mu_0}{\kappa_0}}d\right)\Big/(2\mu) \quad (10a)$$

for $P_x$ and $P_y$ ($i = x, y$), and

$$\lambda_z = \gamma/(2\mu) \quad (10b)$$

for $P_z$. Equation (10a) indicates that the in-plane polarization dynamics is subjected to both the intrinsic damping (described by $\gamma$) and an additional term associated with the radiation electric field $\mathbf{E}^{\mathrm{EM}}$, and that such radiation-induced damping is more significant in thicker films. As shown by Eq. (10b), the out-plane polarization oscillation in the film is not subjected to radiation-induced damping. This is reasonable because $E_z^{\mathrm{EM}}=0$. The angular frequency of oscillation is given as

$$\omega_i = \sqrt{-\frac{A_i+B_i}{\mu}-\lambda_i^2}, \quad (11a)$$

for $P_x$ and $P_y$ ($i = x, y$), and for $P_z$,

$$\omega_z = \sqrt{-\frac{A_z+B_z-\frac{1}{\kappa_0\kappa_b}}{\mu}-\lambda_z^2}, \quad (11b)$$

where,

$$A_i = \frac{\partial E_i^{\mathrm{Landau}}}{\partial P_i}(\mathbf{P}^0)$$

$$= -2\alpha_1-12\alpha_{11}P_i^{02}-2\alpha_{12}\left(P_j^{02}+P_k^{02}\right)$$

$$-30\alpha_{111}P_i^{04}-12\alpha_{112}P_i^{02}\left(P_j^{02}+P_k^{02}\right)-2\alpha_{112}\left(P_j^{04}+P_k^{04}\right)-2\alpha_{123}P_j^{02}P_k^{02}$$

$$-56\alpha_{1111}P_i^{06}-2\alpha_{1112}\left[P_j^{06}+P_k^{06}+15P_i^{04}\left(P_j^{02}+P_k^{02}\right)\right]$$

$$-12\alpha_{1122}P_i^{02}\left(P_j^{04}+P_k^{04}\right)-2\alpha_{1123}(6P_i^{02}P_j^{02}P_k^{02}+P_j^{04}P_k^{02}+P_j^{02}P_k^{04}), \quad (12a)$$

$$B_i = \frac{\partial E_i^{\mathrm{elas}}}{\partial P_i}(\mathbf{P}^0)$$

$$=2[q_{11}(\varepsilon_{ii}-\varepsilon_{ii}^0)+q_{12}(\varepsilon_{jj}+\varepsilon_{kk}-\varepsilon_{jj}^0-\varepsilon_{kk}^0)](\mathbf{P}^0)$$

$$-4(q_{11}Q_{11}+2q_{12}Q_{12})P_i^{0^2}-2q_{44}Q_{44}\left(P_j^{0^2}+P_k^{0^2}\right), \tag{12b}$$

with $i = x, y, z$, and $j \neq i, k \neq i, j$. Based on Eqs. (11-12), we can calculate the intrinsic oscillation frequencies of polarization, i.e., the soft mode phonon frequencies, via $f_i = \frac{\omega_i}{2\pi}$. The relaxation time $\tau_i$ is the time for the polarization variation $\Delta P_i(t)$ to decrease from its peak amplitude $\Delta P_i^0$ to $1/e$ of that value. $\tau_i$ can be calculated as $\tau_i = 1/\lambda_i$, where the expressions of $\lambda_i$ are in Eqs. (10a-b).

Figure 2a and 2b shows the intrinsic frequency $f_i$ and the relaxation time $\tau_i$ of polarization oscillation as a function of the film thickness $d$ in the in-plane-polarized single-domain $(100)_{pc}$ BaTiO$_3$ film, respectively. It is noteworthy that the thickness dependence of both the $f_i$ and $\tau_i$ results from the thickness-dependent radiation electric field $\mathbf{E}^{EM}$. For simplicity, we assume the film is always coherently strained and maintains a single-domain state in all thicknesses. As shown in Fig. 2a, $f_x$ and $f_y$ (the oscillation frequencies of polarization $P_x$ and $P_y$) differ significantly from each other in the $(100)_{pc}$ BaTiO$_3$ film due to the discrepancies in the slope of the local free energy landscape along $P_x$ and $P_y$ (see Figs. 1d and 1e). Both the $f_x$ and $f_y$ decrease as the film thickness $d$ increases due to the more significant radiation-induced damping at larger thicknesses, as mentioned above. Specifically, $f_x$ decreases from 5.59 THz to 0 when $d$ increases from 5 nm to 503 nm, while the $f_y$ decreases from 1.36 THz to 0 when $d$ increases from 5 nm to 122 nm. A zero frequency indicates that the polarization dynamics is critically damped. At larger film thickness, the polarization dynamics is overdamped. There is no polarization oscillation for critically damped or overdamped dynamics, which is analogous to mechanical oscillation. The values of $f_x$ and $f_y$ at $d=5$ nm are almost the same as those calculated without considering $\mathbf{E}^{EM}$ (see dashed lines in Fig. 2a). For completeness, we note that $f_z = 20.66$ THz is thickness-independent (not shown in Fig. 2a) because $E_z^{EM}=0$. The $f_z$ is much higher than the $f_x$ and $f_y$ because of the additional term of $\frac{1}{\kappa_0\kappa_b}$ that results from the depolarization field, and this can be seen by comparing Eqs. (11a) and (11b).

Figure 2a also shows that the influence of $\mathbf{E}^{EM}$ on the intrinsic polarization oscillation frequency is relatively weak in thin ferroelectric films. For example, in a 10-nm-thick film, the $\mathbf{E}^{EM}$ reduces the $f_x$ from 5.590 THz to 5.589 THz, while reduces the $f_y$ from 1.363 THz to 1.357 THz. However, the trend is reverse for the thickness dependence of the oscillation relaxation time $\tau$, which is the same for $P_x$ and $P_y$ (see Eq. 10a). For example, the relaxation time $\tau$ calculated with $\mathbf{E}^{EM}$ is only 1.3 ps in a 10-nm-thick $(100)_{pc}$ BaTiO$_3$ film, which is less than 1/10$^{th}$ of the value (13.5 ps) calculated without $\mathbf{E}^{EM}$, as shown in Fig. 2b. This result demonstrates the necessity of incorporating the radiation electric field to achieve an accurate prediction of ultrafast (THz) polarization dynamics in ferroelectric thin films. Figure 2c shows that mismatch strain can induce large modulation of the $f_x$ and $f_y$ in the in-plane-polarized, 10-nm-thick $(100)_{pc}$ BaTiO$_3$ film. Specifically, when the $\varepsilon_{xx}^{mis}$ increases from 0.01% to 1% ($\varepsilon_{yy}^{mis}$ is fixed at 0.01%), $f_x$ increases from 4.14 THz to 6.7 THz, while $f_y$ increases from 1.04 THz to 1.56 THz. This is consistent with the indication from the slope of the total free energy landscapes (c.f. Fig.1e). Note that the oscillation relaxation time $\tau$ is not influenced by the mismatch strain according to Eq. (10).

## III. Dynamical phase-field simulations

In order to demonstrate the analytically predicted effects of radiation electric field on both the frequency (Fig. 2a) and relaxation time (Fig. 2b) of the soft mode phonons, we performed dynamical phase-field simulations to compute the dynamical oscillation of the polarization in anisotropically strained, single-domain $(100)_{pc}$ BaTiO$_3$ film following the excitation of a single ultrafast THz electric-field pulse. Compared to existing dynamical phase-field models that only consider coupled polarization and strain dynamics [35,36], our model considers the coupled dynamics of strain, polarization, and EM waves by solving the nonlinear equations of motions for the polarization **P** (see Eq. 1), mechanical displacement **u**, and the radiation electric field $\mathbf{E}^{EM}$ in a coupled fashion. Specifically, the spatiotemporal distribution of the $\mathbf{E}^{EM}$ and associated radiation magnetic field $\mathbf{H}^{EM}$ are obtained by solving the Maxwell's equations,

$$\frac{\partial \mathbf{E}^{EM}}{\partial t} = \frac{1}{\kappa_0 \kappa_b}(\nabla \times \mathbf{H}^{EM} - \frac{\partial \mathbf{P}}{\partial t}) \tag{13a}$$

$$\frac{\partial \mathbf{H}^{EM}}{\partial t} = -\frac{1}{\mu_0}\nabla \times \mathbf{E}^{EM} \tag{13b}$$

where the term $\frac{\partial \mathbf{P}}{\partial t}$ describes the time-varying polarization as the source of electric dipole radiation. The polarization and strain dynamics are coupled since the elastic effective electric field $\mathbf{E}^{elas}$ is a function of the polarization **P**, stress-free strain $\boldsymbol{\varepsilon}^0$, and total strain $\boldsymbol{\varepsilon}$, as shown in Eq. (3). Here, the total strain is calculated as the sum of the component at the initial equilibrium state and the temporal variation component, that is, $\boldsymbol{\varepsilon}=\boldsymbol{\varepsilon}(\mathbf{P}^0)+\Delta\boldsymbol{\varepsilon}(t)$. The expression of the $\boldsymbol{\varepsilon}(\mathbf{P}^0)$ is given by Eq. (5). The $\Delta\boldsymbol{\varepsilon}(t)$ is calculated as $\Delta\varepsilon_{ij}=\frac{1}{2}\left(\frac{\partial \Delta u_i}{\partial j}+\frac{\partial \Delta u_j}{\partial i}\right)$, where $\Delta\mathbf{u}(t)$ is the temporal variation of the mechanical displacement **u** with $i, j = x, y, z$. The $\Delta\mathbf{u}(t)$ is obtained by solving the elastodynamics equation for the entire film-on-substrate heterostructure,

$$\rho\frac{\partial^2 \Delta\mathbf{u}}{\partial t^2} = \nabla \cdot [\mathbf{c}(\Delta\boldsymbol{\varepsilon}-\Delta\boldsymbol{\varepsilon}^0)] \tag{14}$$

where $\rho$ and **c** are phase-dependent mass density and elastic stiffness tensor, respectively; $\Delta\boldsymbol{\varepsilon}^0(t) = \boldsymbol{\varepsilon}^0(t)-\boldsymbol{\varepsilon}^0(\mathbf{P}^0)$ is the temporal variation of the stress-free strain. Free surface mechanical boundary condition $\sigma_{iz}=0$ ($i = x, y, z$) is applied on the top surface of the ferroelectric film during the dynamics. The entire heterostructure is discretized into one-dimensional computational cells along $z$ axis with the cell size being 1 nm. The $(110)_O$ PrScO$_3$ substrate and the free space above the ferroelectric film are discretized into 100 and 10 cells, respectively. Absorbing boundary condition (ABC) is applied on both bottom and top surfaces of the computational system when solving the Maxwell's equations for the generation and propagation of EM waves. The ABC is also applied at the bottom surface of the substrate for making the substrate a perfect sink for elastic waves. Central finite difference is used to calculate spatial derivatives in all dynamical equations (Eqs. 1, 13 and 14). And these equations are solved simultaneously in a coupled manner using classical Runge-Kutta method for time marching with a time step of $10^{-18}$ s. Specifically, the Maxwell's equations are solved using classical finite-difference time-domain (FDTD) method. Relevant materials parameters for the BaTiO$_3$ film and PrScO$_3$ substrate are provide in Supplementary Material 3. Moreover, our in-house numerical solvers for all equations are GPU-accelerated where the computations for each simulation cell are performed parallelly.

To numerically demonstrate the analytical prediction for in-plane-polarized (100)$_{pc}$ BaTiO$_3$ film, harmonic oscillation of both the $P_x$ and $P_y$ component needs to be excited. Here, this is achieved by applying the THz electric-field pulse $\mathbf{E}^{app}(t)$ along the $y$ axis. The waveform of $\mathbf{E}^{app}(t)$, as shown in the inset of Fig. 3a, takes the form of $E_y^{app}(t)=E_0^{app}\exp\left[-\frac{(t-t_0)^2}{w^2}\right]\cos[\omega^{app}(t-t_0)]$ following ref. [18]. Here $\omega^{app}/2\pi =1.5$ THz defines the peak frequency of the electric-field pulse. $E_0^{app} = 10^6$ V m$^{-1}$ and $t_0 = 0.6$ ps are the amplitude and temporal position of the peak electric field, respectively, while $w = 0.2$ ps is the width of Gaussian function. The selection of the values for these parameters follows two criteria. First, the electric-field pulse cannot be too strong and/or long, otherwise the polarization dynamics will be driven into the anharmonic region. Second, the frequency spectrum of THz electric-field pulse needs to contain the frequencies of the soft mode phonons. Specifically, the $\Delta\mathbf{P}$ will have larger amplitude in real-space when the THz pulse has a larger spectral amplitude at the soft mode phonon frequencies (see Supplementary Material 4). At the initial equilibrium state, the polarization in the 10-nm-thick single-domain (100)$_{pc}$ BaTiO$_3$ film is along the $+x$-axis. Once a nonzero $P_y$ is induced by $E_y^{app}$, a nonzero $\Delta P_x(t)=[P_x(t)-P_x^0]$ will also be developed due to the nonzero but much smaller $E_x^{Landau}$ (see Eq. 2). As a result, both the $E_y^{EM}$ and $E_x^{EM}$ are nonzero, but the amplitude of the $E_y^{EM}$ is two orders of magnitude larger, as shown in Fig. 3a and 3b. Of course, one can also apply the electric-field pulse along the in-plane diagonal axis, in which case $P_x$ and $P_y$ can be directly excited by the nonzero $E_x^{app}$ and $E_y^{app}$, respectively.

Figure 3c shows the evolution of the $\Delta P_y(t)$ in a 10-nm-thick (100)$_{pc}$ BTO film with and without considering the radiation electric field $E_y^{EM}$. Note that $\Delta P_y(t)=P_y(t)$ since $P_y^0=0$. As shown, the $P_y$ reaches its peak amplitude at ~1.1 ps and decreases monotonically due to the damping. According to the analytical prediction, the $E_y^{EM}$ reduces the intrinsic oscillation frequency of $P_y$ from 1.363 THz to 1.357 THz (Fig. 2a) and the relaxation time from 13.5 ps to 1.3 ps (Fig. 2b) due to the enhanced damping. The reduced relaxation time can be clearly seen from Fig. 3c. Notably, the decrease in the peak values of $\Delta P_y$ at each oscillation period can be well described by the analytical expression $|\Delta P_y|(t)=|\Delta P_y^0|e^{-\frac{(t-t^0)}{\tau}}$, where $\tau$ is the analytically calculated relaxation time (13.5 ps or 1.3 ps) and $\Delta P_y^0=P_y(t = 1.1$ ps$)$ is the peak amplitude (indicated in Fig. 3c). The analytically calculated decreasing trajectories are shown as the dashed lines. The remarkable agreement demonstrates that the analytically calculated $\tau$ is accurate. Figure 3e shows the frequency spectra of the simulated $\Delta P_y(t)$ curves with and without $E_y^{EM}$ (see red and black curves). Both curves show a single peak at the 1.36 THz, which is consistent with analytical prediction but the predicted frequency shift from 1.363 THz to 1.357 THz (0.006 THz discrepancy) is too small to resolve numerically. However, our simulations using thicker BTO film and adjusted materials parameters clearly confirms the analytically predicted frequency shift (see Supplementary Material 5). For completeness, the frequency spectrum of the $E_y^{EM}(t)$ is also plotted in Fig. 3e, which displays a single peak at 1.36 THz as expected.

The numerically simulated $\Delta P_x(t)$ in the 10-nm-thick (100)$_{pc}$ BTO film is analyzed in a similar manner to $\Delta P_y(t)$. The results are summarized in Fig. 3d and 3f. One notable difference is that the $P_x(t)$ and $E_x^{EM}(t)$ have two peak frequencies at 2.72 THz and 5.6 THz. The peak of 5.6 THz is consistent with the analytically calculated intrinsic frequency of the $P_x$ (Fig. 2a). Likewise, the analytically predicted frequency shift from 5.59 THz (without $E_x^{EM}$) to 5.589 THz (with $E_x^{EM}$) is too small to resolve with the present numerical accuracy. The peak of 2.72 THz, which is exactly twice the intrinsic frequency of the $P_y$ (1.36 THz), results from the second-order harmonic of the $P_y$ oscillation, which is described by the terms containing $P_y^2$ in the expression of $E_x^{\text{Landau}}$ (Eq. 2). From the frequency spectra in Fig. 3f, it can be seen that the second order harmonic component (2.72 THz) associated with $P_y^2$ outweighs the intrinsic component (5.6 THz) associated with the oscillating $P_x$. In view of this, we use a modified analytical formula $|\Delta P_x|(t)=|\Delta P_x^0|e^{-\frac{2(t-t^0)}{\tau}}$ to predict the decrease in the peak values of $\Delta P_x$ at each oscillation period, where $\tau$ is again the analytically calculated relaxation time (13.5 ps or 1.3 ps) and $\Delta P_x^0=P_x(t=1.1 \text{ ps})$ is the peak amplitude (indicated in Fig. 3d). As shown by the dashed lines in Fig. 3d, the analytical calculations are mostly consistent with numerical simulations with discrepancies arising due to the simultaneous presence of the 5.6 THz component. Likewise, the frequency spectrum of the radiation electric field $E_x^{\text{EM}}$ is plotted in Fig. 3f for comparison, which also displays two distinct peaks at 2.72 THz and 5.6 THz, respectively. Furthermore, it is worth remarking that both the $E_x^{\text{EM}}$ and $E_y^{\text{EM}}$ have a sufficiently large amplitude in the time-domain (Figs. 3a-b) for near-field detection. In this regard, the predicted THz-pulse-induced polarization oscillation in single-domain $(100)_{\text{pc}}$ BaTiO$_3$ thin film on $(110)_{\text{O}}$ PrScO$_3$ substrate can in principle be experimentally validated by THz-pump THz-probe spectroscopy, through which both the soft mode phonon frequencies and relaxation time can be experimentally determined.

## IV. Conclusions

In summary, our results show that an accurate modeling of ultrafast THz-field excitation of ferroelectrics requires considering the polarization-photon coupling. Using ultrafast THz-field-induced harmonic polarization oscillation in single-domain ferroelectric thin film as an example, we analytically and numerically demonstrate that the polarization-oscillation-induced radiation electric field can reduce the frequency and relaxation time of soft mode phonons by increasing the effective damping. We have derived the analytical expressions on the variation of the soft mode phonon frequencies as a function of epitaxial strain and thin film thickness, which can be utilized to guide the synthesis of thin films and shape engineering of THz pulse for achieving more efficient resonant excitation. The soft mode phonon (i.e., intrinsic polarization oscillation) frequency is analogous to ferromagnetic resonance (FMR) frequency in ferromagnetic system, which describes the intrinsic precession frequency of uniform magnetization and can likewise be modulated by epitaxial strain [37]. However, in contrast with the strong influence of the radiation electric field on polarization dynamics demonstrated herein, the radiation magnetic field produced by the uniform magnetization precession is negligible in nanometer-thick ferromagnetic thin films and hence barely influence the FMR frequency and effective magnetic damping. Furthermore, our GPU-accelerated dynamical phase-field model, which considers fully coupled dynamics of strain, polarization, and EM waves, can be extended to model polarization dynamics in more complex ferroelectric materials under the excitation by other types of ultrafast light pulses.

**Figures and Captions**

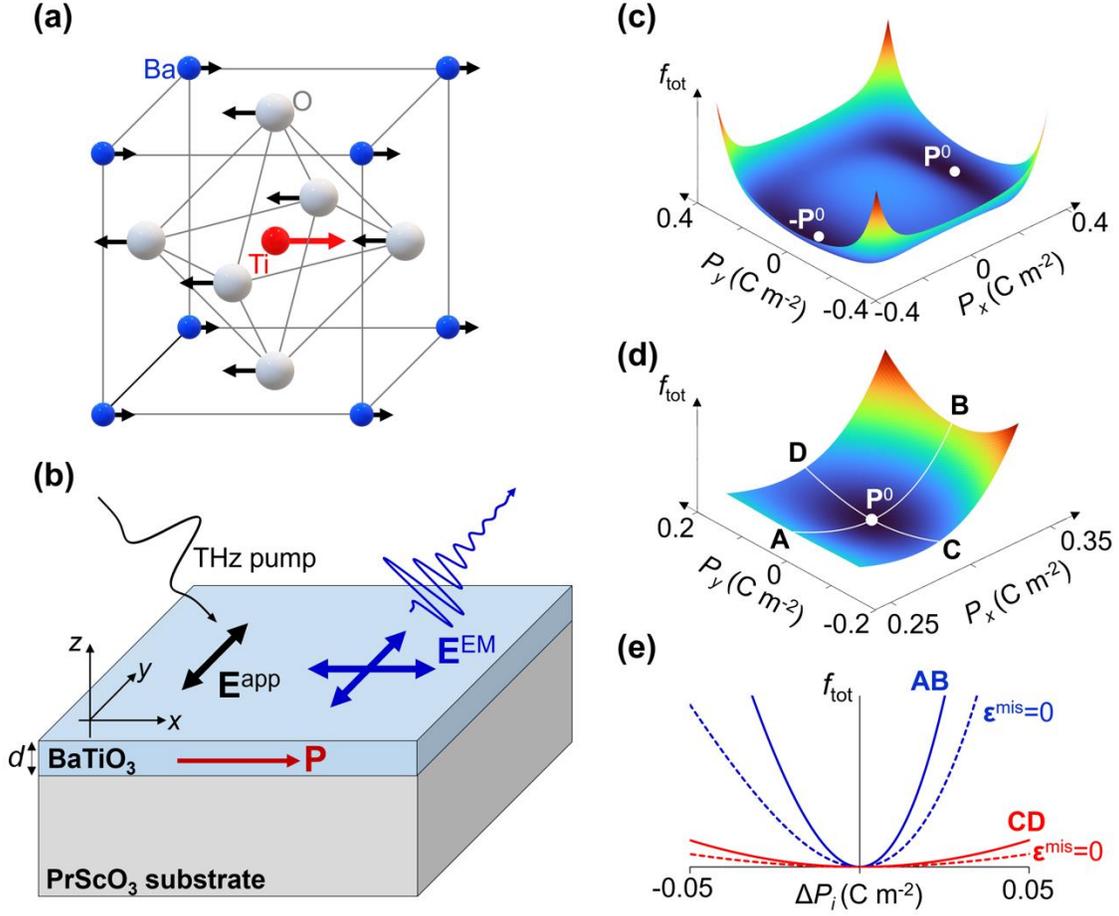

**Figure 1**. (**a**) Schematic of the soft mode phonon $A_1$(TO) in $BaTiO_3$. (**b**) Excitation of the soft mode phonons in the $BaTiO_3$(film)/$PrScO_3$(substrate) heterostructure by THz electric-field pulse $\mathbf{E}^{app}$. The radiation electric field $\mathbf{E}^{EM}$ from the THz-excited oscillating polarization $\mathbf{P}$ will in turn influence the polarization dynamics. (**c**) Total free energy landscape for anisotropically strained single-domain $(100)_{pc}$ $BaTiO_3$ film as a function of in-plane polarization components $P_x$ and $P_y$. The $\mathbf{P}$ locates at local free energy minima at equilibrium, which are labeled as $\pm\mathbf{P}^0$. (**d**) Enlarged total free energy landscape near the $+\mathbf{P}^0$. Two lines in the free energy that cross the $+\mathbf{P}^0$ and vary long $P_x$ and $P_y$ (white solid lines) are labeled as AB and CD, respectively. (**e**) 2D plot of the data in line AB and CD, and their counterparts under zero mismatch strain. $\Delta P_i = P_i - P_i^0$, $i = x, y$, where the $P_i^0$ calculated with and without the mismatch strain are different (details are in text).

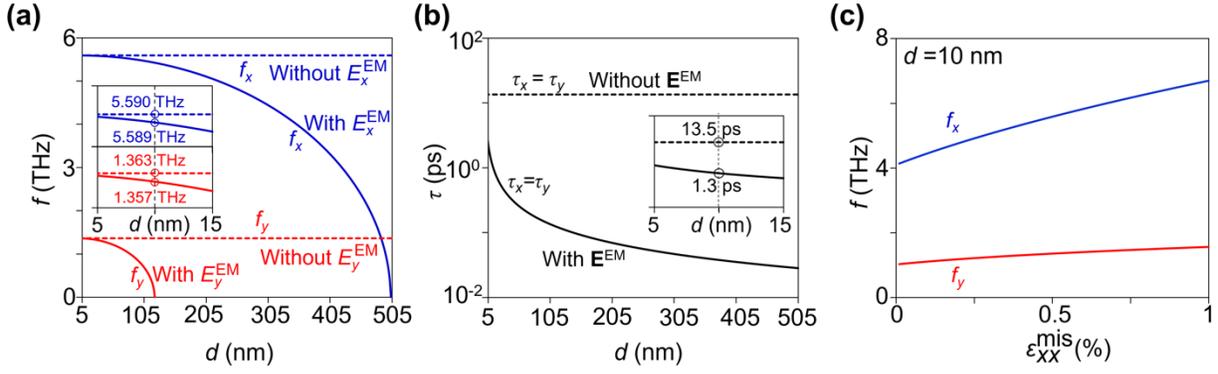

**Figure 2.** (a) Analytically calculated $f_x$ and $f_y$ (intrinsic oscillation frequency of $P_x$ and $P_y$) as a function of the $(100)_{pc}$ BaTiO$_3$ film thickness $d$, with (solid lines) and without (dashed lines) considering the $\mathbf{E}^{EM}$. The enlarged $f$-$d$ curves from $d = 5$ nm to 15 nm are shown in the inset and the frequency values for $d = 10$ nm are labeled. (b) Analytically calculated relaxation time $\tau$ of $P_x$ and $P_y$ oscillation as a function of $d$, with (solid line) and without (dashed line) considering the $\mathbf{E}^{EM}$. The enlarged $\tau$-$d$ curves from $d = 5$ nm to 15 nm are shown in the inset and the $\tau$ values for $d = 10$ nm is labeled. (c) Analytically calculated $f_x$ and $f_y$ as a function of the mismatch strain component $\varepsilon_{xx}^{mis}$ at $d$=10 nm. $\varepsilon_{yy}^{mis}$ is fixed as 0.01% and $\varepsilon_{xy}^{mis}$=0. $\mathbf{E}^{EM}$ is considered.

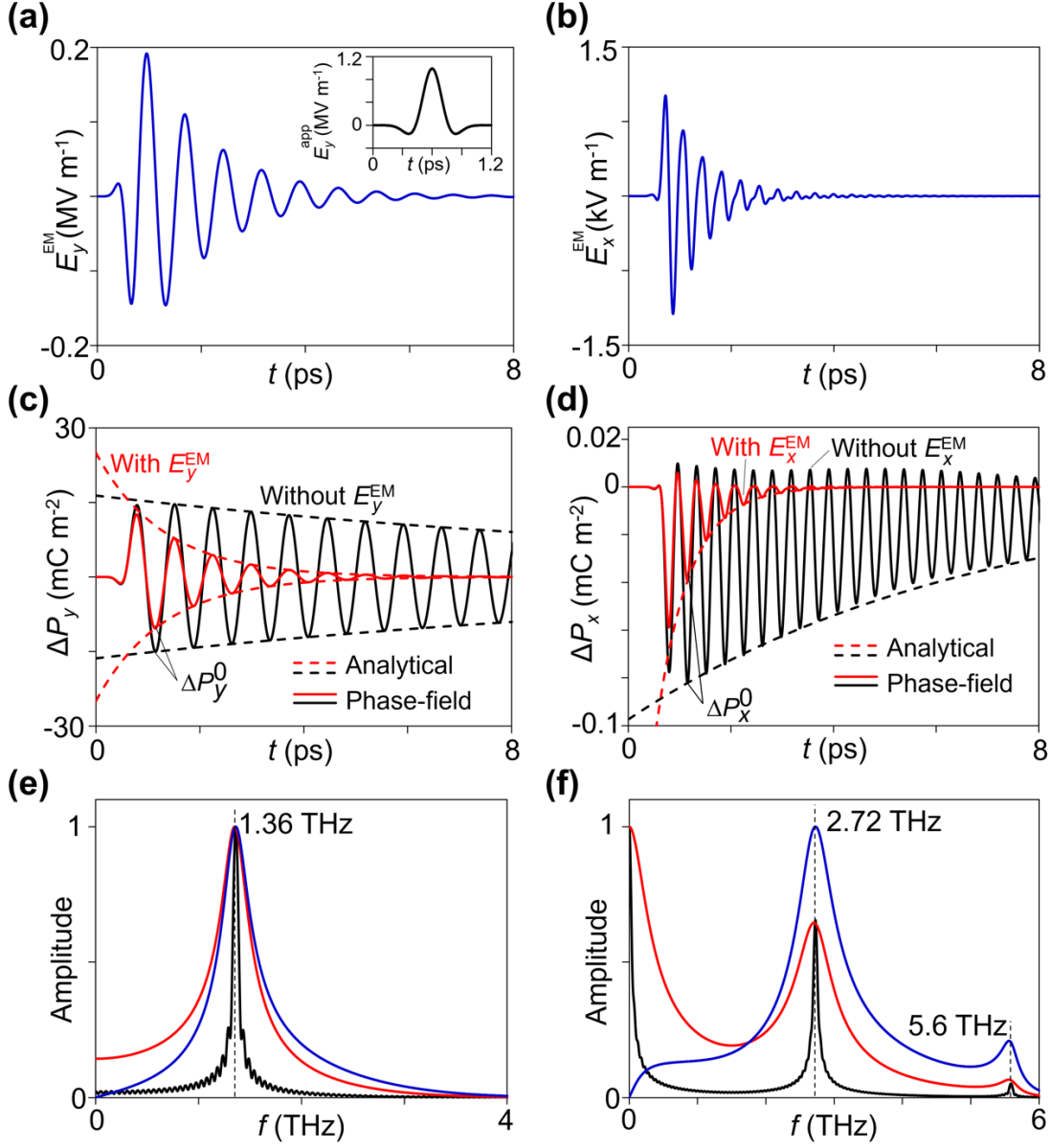

**Figure 3**. Profiles of the numerically simulated **(a)** $E_y^{EM}(t)$ and **(b)** $E_x^{EM}(t)$ at 2 nm above the $(100)_{pc}$ BaTiO$_3$ film upon the excitation by a single ultrafast THz electric-field pulse $\mathbf{E}^{app}$, which is applied along the *y*-axis and its temporal profile is shown in the inset of (a). Simulated oscillation trajectories of **(c)** $\Delta P_y(t)$ and **(d)** $\Delta P_x(t)$ with (red solid line) and without (black solid line) considering their conjugate $\mathbf{E}^{EM}$ component. The analytically calculated decrease in the peak values of $\Delta P_i$ ($i = x, y$) are plotted using the dashed lines. **(e)** Frequency spectra of the $E_y^{EM}(t)$ in (a) and the two solid lines of $\Delta P_y(t)$ in (c). **(f)** Frequency spectra of the $E_x^{EM}(t)$ in (b) and the two solid lines of $\Delta P_x(t)$ in (d). The peak frequencies are labeled.

**Acknowledgements**

J.-M.H. acknowledges support from the NSF award CBET-2006028 and the Accelerator Program from the Wisconsin Alumni Research Foundation. The simulations were performed using Bridges at the Pittsburgh Supercomputing Center through allocation TG-DMR180076, which is part of the Extreme Science and Engineering Discovery Environment (XSEDE) and supported by NSF grant ACI-1548562.